\documentclass[prd,superscriptaddress,onecolumn,showpacs,amsmath,amssymb]{revtex4}
\usepackage{graphicx}
\usepackage{color}
\usepackage{dcolumn}
\usepackage{bm}
\usepackage{slashed}

\begin{document}
\title{Relationship between Bondi-Sachs quantities and source of gravitational radiation in asymptotically de Sitter spacetime}

\author{Xiaokai He}
\email[Xiaokai He: ]{hexiaokai77@163.com} \affiliation{Department of Physics, Key
Laboratory of Low Dimensional Quantum Structures and Quantum Control
of Ministry of Education, and Synergetic Innovation Center for
Quantum Effects and Applications, Hunan Normal University, Changsha,
Hunan 410081, P. R. China}

 \affiliation{School of Mathematics and
Computational Science, Hunan First Normal University, Changsha
410205, China}

\author{Jiliang Jing}
\email[Jiliang Jing: ]{jljing@hunn.edu.cn} \affiliation{Department of Physics, Key
Laboratory of Low Dimensional Quantum Structures and Quantum Control
of Ministry of Education, and Synergetic Innovation Center for
Quantum Effects and Applications, Hunan Normal University, Changsha,
Hunan 410081, P. R. China}

\author{Zhoujian Cao}
\email[Zhoujian Cao: ]{zjcao@amt.ac.cn} \affiliation{Department of Astronomy, Beijing Normal University, Beijing 100875, China}
\affiliation{Academy of Mathematics and Systems Science,
Chinese Academy of Sciences, Beijing 100190, China}

\begin{abstract}
Gravitational radiation  plays an important role in astrophysics.
Based on the fact that our universe is expanding, the gravitational
radiation when a positive cosmological constant is presented has
been studied along with two different ways recently, one is the
Bondi-Sachs (BS) framework in which the result is shown by  BS quantities
in the asymptotic null structure, the other is the perturbation approach in
which the result is presented by the quadrupoles of source.  Therefore, it is
worth to interpret the quantities in asymptotic null structure in terms of the information of the source. In this paper, we investigate this problem and find the  explicit expressions of BS quantities in terms of the quadrupoles of source in asymptotically de Sitter spacetime.  We also estimate how far away
the source is, the cosmological constant may affect the detection of the gravitational wave.
\end{abstract}


\maketitle

\section{Introduction}

Almost one hundred years ago, Einstein provided a relativistic
description of gravitational wave by linearizing the field equation
of general relativity (GR). Moreover he calculated the energy
carried away by gravitational radiations emitted by a varying mass
quadruple moment in harmonic gauge condition \cite{einstein18}.
However, due to the complexity of the diffeomorphism of GR, even the
existence of gravitational waves in  GR caused a great controversy
for a long time \cite{Kennefick07}.  This controversy was finally
resolved theoretically in 1960s by the work of Bondi, Sachs and
others \cite{BonVanMet62,Sac62}. In their work, they constructed
Bondi-Sachs (BS) framework and obtained the mass loss formula in
terms of the so-called ``news function".

It should be noted that Einstein's calculations, as well as the BS
framework in  GR, used the field equations without cosmological
constant. In the past years, there are strong evidences
\cite{deAdeBoc00,KomSmiDun11,PerAldDel98,RieFilCha98,Pla13,Pad03}
from many independent astronomical observations show that our
present universe is in a state of accelerating expansion and is best
modeled by cold dark matter with a positive cosmological constant
$\Lambda$. Then it is worth to recheck the behavior of gravitational
wave based on the Einstein's equation with a cosmological constant
\begin{eqnarray}
R_{\mu\nu}-\frac{1}{2}Rg_{\mu\nu}+\Lambda g_{\mu\nu}=8\pi T_{\mu\nu},
\end{eqnarray}
where we have used geometric units with $c=G=1$.

Very recently, three gravitational wave detection events have been confirmed \cite{Abbott16,Abbott16(2),PhysRevLett.118.221101}.
The sources of these two events locate about billion light years away which is not very far away compared to the
cosmological scale. The experiment results show that the theory of Einstein, Bondi and others \cite{einstein18,BonVanMet62,Sac62}
works quite well. No sign of cosmological effect on gravitational wave shows up. In the near future
when the sensitivity of advanced LIGO and other ground based detectors grows more, farther
gravitational wave events will be detected and the  effect of cosmological constant will become
more and more significant. Surprisingly, Ashtekar and his coworkers showed that
the cosmological constant $\Lambda$ may change the behavior of gravitational wave
strongly no matter how small it is \cite{Ashtekar16}. They found theoretically that
gravitational wave does not exist for such far away sources if the spacetime is strongly asymptotical
de Sitter \cite{Ashtekar2015I,Ashtekar2015II,Ashtekar16}.
Later, they used perturbation method based on the work of de Vega \cite{Vega1999} to find that the
gravitational wave may exist also and gave the quadrupole formula when the cosmological constant presents. In Refs.~\cite{HeCao15,HeCaoJing16}, we analyzed the asymptotic behavior of spacetime with a cosmological constant by the BS framework. By proposing a new Bondi-type out going boundary condition for Einstein equation , we reconcile the theoretical tension between the cosmological constant and the existence of gravitational wave. More explicitly, we calculated the Newman-Penrose scalar $\Psi_4$ which reduce to Bondi's original result when $\Lambda$ goes to zero and found that when gravitational radiation was emitted from the isolated source, the conformal boundary of spacetime with cosmological constant is not conformally flat, i.e., the spacetime is only weakly asymptotically de Sitter. In Ref.~\cite{Saw16}, Saw discussed our new boundary condition based on Newman-Penrose formalism. Recently, Bishop use BS framework to investigate the
solution of linearized Einstein equations with a cosmological constant and showed that gravitational wave energy conservation does not normally apply to inertial observers \cite{Bishop16}.

It is worth to interpret the asymptotic null structure in
terms of the information of the source. For the case
of Einstein equation without cosmological constant, Coleman
investigated the relations between harmonic and BS coordinates by
solving wave equations in 1974 and found an approximation to the
news function  in terms of the energy momentum tensor of the source
\cite{Coleman}. But for the case of Einstein equation with a
cosmological constant, there is little results yet about the
relationship between the gravitational wave ``news" in BS framework
and the source behavior.  In this work, we intend to investigate
this problem  based on the result of Ref.~\cite{Date2015} and discuss the
outgoing boundary condition in the presence of a cosmological
constant. We found that Coleman's method is hard to be generalized
to the case including a cosmological constant. Different to
Coleman's method which transforms the BS coordinate to harmonic
coordinate, we transform instead the harmonic coordinate to BS coordinate
and finally express quantities appeared in our new outgoing boundary
condition and gravitational wave ``news" in terms of the quadrupole
of the source.

The rest of the paper is organized as follows. In the next section,
we will solve the linearized Einstein equation with a cosmological
constant for bounded gravitational wave sources. In Sec.~III, we
transform the solution got in Sec.~II to the transverse-traceless
gauge. After that we find out the Bondi-Sachs coordinate based on
the transverse-traceless gauge in Sec.~IV. Based on the new Bondi-type
out going boundary condition, in Sec.~V we review the result for
Einstein equation together with a cosmological constant for
axisymmetric spacetime and more extending our previous result to
general spacetime case. Along the description in Sec.~V, we give the
expression of BS quantities in terms of the quadrupole of the
gravitational wave source and  discuss  our new Bondi-type out going
boundary condition. In section VI, we calculate the gravitational
news function for an equal mass binary system based on the result obtained in
Sec.~V. At
last, we summarize this paper in Sec.~VII with some comments and
discussions.

Throughout this paper, we use the signature convention $(-,+,+,+)$
for spacetime metric and geometric units with $G=c=1$. Moreover, the
Einstein summation convention is also adopted. For the indexes, the Latin
indices are spatial indices and run from 1 to 3, whereas
Greek indices are space-time indices and run from 0 to 3. Following the convention of Ref.~\cite{liang00}, we use the notation `$:=$' to denote the definition and the notation `$\equiv$' to mean an identity.

\section{gravitational wave solution to the linearized Einstein equation with a cosmological constant}

The Minkowski spacetime is a solution to the Einstein equation without cosmological constant for empty space. And the Minkowski spacetime is a maximum symmetric spacetime. Similarly, the de Sitter spacetime is a solution to the Einstein equation with a cosmological constant for empty space. And it is also a maximum symmetric spacetime. So when the gravitational wave source is weak, it is natural to consider the perturbation of de Sitter spacetime when a positive cosmological constant is presented.

When a small perturbation of de Sitter spacetime was considered, the
metric becomes \cite{Ashtekar2015I,Ashtekar2015III,Date2015}
\begin{eqnarray}
g_{\mu\nu}=\gamma_{\mu\nu}+\epsilon h_{\mu\nu},
\end{eqnarray}
where $\gamma_{\mu\nu}$ is the de Sitter metric and $\epsilon$ is a
smallness parameter. Similar to the Minkowski perturbation,
it is convenient to introduce the trace-reversed perturbation
\begin{eqnarray}
\bar{h}_{\mu\nu}=h_{\mu\nu}-\frac{1}{2}\gamma_{\mu\nu}h,
\end{eqnarray}
where $h$ is the trace of the perturbation $h\equiv\gamma^{\mu\nu}h_{\mu\nu}$.
It is simpler to solve the linearized Einstein equation based on the conformal
coordinate $(\eta,x,y,z)$ in which the de Sitter metric is written as
\begin{eqnarray}
ds^2=\frac{1}{H^2\eta^2}[-d\eta^2+dx^2+dy^2+dz^2],
\end{eqnarray}
where $H\equiv\sqrt{\frac{\Lambda}{3}}$. By imposing more the generalized Lorentz/harmonic gauge condition
\begin{eqnarray}
\nabla^{\mu}\bar{h}_{\mu\nu}=2H^2\eta\bar{h}_{0\nu},\label{GLC}
\end{eqnarray}
where $\nabla^{\mu}$ is the covariant derivative operator respect to $\gamma_{\mu\nu}$,
the linearized Einstein equation reduces to \cite{Date2015}
\begin{align}
&\eta^{\alpha\beta}\partial_\alpha\partial_\beta
\chi_{\mu\nu}+\frac{2}{\eta^2}\partial_0\chi_{\mu\nu}\nonumber\\
&-\frac{2}{\eta^2}\left(\delta^0{}_{\mu}\delta^0{}_{\nu}\eta^{\alpha\beta}\chi_{\alpha\beta}
+\delta^0{}_{\mu}\chi_{0\nu}+\delta^0{}_{\nu}\chi_{0\mu}\right)=-16\pi
T_{\mu\nu},\label{lineareinstein}
\end{align}
where $\eta_{\alpha\beta}$ is the Minkowski metric.
Here we have replaced the variable $\bar{h}_{\mu\nu}$ with
\begin{eqnarray}
\chi_{\mu\nu}:=H^2\eta^2\bar{h}_{\mu\nu}.
\end{eqnarray}
As shown in Ref.~\cite{Date2015}, in addition to the Lorenz/harmonic gauge condition (\ref{GLC}) synchronous gauge condition can be used to make
\begin{align}
\bar{h}_{0\mu}=0=\chi_{0\mu}.\label{SGC}
\end{align}
Regarding to the $\chi_{ij}$ part,
based on the Hadamard ansatz, the solution to the Eq.~(\ref{lineareinstein}) can be written as
\begin{eqnarray}
\chi_{ij}(\eta,\vec{x})=4\int
d^3\vec{x}\frac{1}{|\vec{x}-\vec{x}'|}T_{ij}(\eta-|\vec{x}-\vec{x}'|,\vec{x}')\nonumber\\
+
4\int
d^3\vec{x}'\int_{-\infty}^{\eta-|\vec{x}-\vec{x}'|}\frac{d\eta'}{\eta'}\frac{\partial
T_{ij}(\eta',\vec{x}')}{\partial\eta'},
\end{eqnarray}
where $\vec{x}\equiv(x,y,z)$ and $|\vec{x}|\equiv\sqrt{x^2+y^2+z^2}$. The integration is respect to the gravitational wave source region which is denoted with coordinate $\vec{x}'$.
In the wave zone where $|\vec{x}|>>|\vec{x}'|$, we can use the
approximation $|\vec{x}-\vec{x}'|\simeq R := |\vec{x}|$. Then we can get
\begin{eqnarray}
\chi_{ij}(\eta,\vec{x})=\frac{4\eta}{r(\eta-R)}\int
d^3\vec{x}'T_{ij}(\eta-R,\vec{x}')\nonumber\\
-4\lim\limits_{\eta'\rightarrow -\infty}\frac{1}{\eta'}\int
d^3\vec{x}'T_{ij}(\eta',\vec{x}')\nonumber\\
+4\int
d^3x'\int_{-\infty}^{\eta-R}d\eta'\frac{T_{ij}(\eta',x')}{\eta'^2}\label{linearsolution}
\end{eqnarray}
via a partial integration.

In order to relate the above solution to the
BS framework, it is simpler to express the solution in the cosmological coordinate $(t,x,y,z)$ in
which the de Sitter metric can be written as
\begin{eqnarray}
ds^2=-dt^2+e^{2Ht}[dx^2+dy^2+dz^2].
\end{eqnarray}
The coordinate $x$, $y$, $z$ are the same to the ones in the above conformal
coordinate while the coordinate $t$ is related to the above $\eta$ through
\begin{eqnarray}
\eta=-\frac{1}{H}e^{-Ht}.
\end{eqnarray}
Based on the cosmological coordinate, by using the matter
conservation equations, the solution (\ref{linearsolution}) can be expressed as
\begin{align}
\chi_{ij}=&\frac{2}{R}(e^{-Ht}+HR)\bigg{[}\ddot{Q}_{ij}-2H\dot{Q}_{ij}+H\dot{\bar{Q}}_{ij}\bigg{]}\nonumber\\
&-2H\bigg{[}\ddot{Q}_{ij}-3H\dot{Q}_{ij}+H\dot{\bar{Q}}_{ij}+2H^2Q_{ij}-H^2\bar{Q}_{ij}\bigg{]},\label{chi}
\end{align}
where $R\equiv\sqrt{x^2+y^2+z^2}$. Here we have used the notation $Q_{ij}$ for the
 mass quadrupole moment and $\bar{Q}_{ij}$ for the pressure quadrupole moment
\begin{eqnarray}
Q^{ij}=\int_{source}d^3\vec{x}\sqrt{p}\rho x^ix^j,\\
\bar{Q}^{ij}=\int_{source}d^3\vec{x}\sqrt{p}\pi x^ix^j,
\end{eqnarray}
where $p$ is the determinant of the three metric $p_{\mu\nu}$ respect to the
constant $t$ slice, $\rho\equiv T_{\mu\nu}n^\mu n^\nu$ with $n^\mu$ the future
directed normal vector respect to the constant $t$ slice, and $\pi\equiv T_{\mu\nu}p^{\mu\nu}$.
For lower index $Q_{ij}$ and $\bar{Q}_{ij}$ we used $p_{ij}$ to lower the indexes.

It is easy to see that the solution $(\ref{chi})$ recovers the
solution to the Einstein equation without cosmological constant \cite{einstein18} as the cosmological constant goes to
zero.

\section{reexpress the solution to linearized Einstein equation with a cosmological constant using transverse traceless gauge}

According to the discussion in the above section, the solution
to the linearized Einstein equation around de Sitter spacetime can be written as
\begin{eqnarray}
ds^2=-dt^2+e^{2Ht}(\delta_{ij}+\epsilon
\chi_{ij})dx^idx^j\label{desitter1}
\end{eqnarray}
where $\chi_{ij}$ is given by Eq.~$(\ref{chi})$.

In order to avoid the caustics of null curves, we consider Bondi-Sachs framework within the spacetime region out of some cylinder. So it is convenient to consider a spherical coordinate $(R,\theta,\phi)$ respect to the Cartesian coordinate $(x,y,z)$ above
\begin{align}
x&=R\sin\theta\cos\phi,\nonumber\\
y&=R\sin\theta\sin\phi,\nonumber\\
z&=R\cos\theta.\nonumber
\end{align}
Then the metric (\ref{desitter1}) can be described in $(t,R,\theta,\phi)$ as
\begin{align}
ds^2=&-dt^2+e^{2Ht}\bigg{[}(1+\epsilon\chi_{\hat{R}\hat{R}})dR^2+2\epsilon
R\chi_{\hat{R}\hat{\theta}}dRd\theta\nonumber\\
&+2 \epsilon R\sin\theta\chi_{\hat{R}\hat{\phi}}dRd\phi+R^2(1+\epsilon\chi_{\hat{\theta}\hat{\theta}})d\theta^2\nonumber\\
&+2\epsilon R^2\sin\theta\chi_{\hat{\theta}\hat{\phi}} d\theta d\phi+
R^2\sin^2\theta(1+\epsilon\chi_{\hat{\phi}\hat{\phi}})d\phi^2
\bigg{]},\label{desitter2}
\end{align}
where
\begin{align}
\chi_{\hat{R}\hat{R}}=&\chi_{33}\cos^2\theta+\chi_{11}\cos^2\phi\sin^2\theta+\chi_{13}\cos\phi\sin2\theta\nonumber\\
&+\chi_{23}\sin2\theta\sin\phi+\chi_{22}\sin^2\theta\sin^2\phi\nonumber\\
&+\chi_{12}\sin^2\theta\sin
2\phi,\nonumber\\
\chi_{\hat{R}\hat{\theta}}=&(\chi_{13}\cos\phi+\chi_{23}\sin\phi)\cos^2\theta\nonumber\\
&-(\chi_{13}\cos\phi+\chi_{23}\sin\phi)\sin^2\theta\nonumber\\
&+(-\chi_{33}+\chi_{11}\cos^2\phi+\chi_{22}\sin^2\phi\nonumber\\
&+\chi_{12}\sin2\phi)\cos\theta\sin\theta,\nonumber\\
\chi_{\hat{R}\hat{\phi}}=&(\chi_{23}\cos\phi-\chi_{13}\sin\phi)\cos\theta\nonumber\\
&+\frac{1}{2}\sin\theta(2\chi_{12}\cos2\phi+\chi_{22}\sin2\phi-\chi_{11}\sin2\phi),\nonumber\\
\chi_{\hat{\theta}\hat{\theta}}=&\chi_{33}\sin^2\theta-\sin2\theta(\chi_{13}\cos\phi+\chi_{23}\sin\phi)\nonumber\\
&+\cos^2\theta(\chi_{11}\cos^2\phi+\chi_{22}\sin^2\phi+\chi_{12}\sin2\phi),\nonumber\\
\chi_{\hat{\theta}\hat{\phi}}=&\sin\theta(-\chi_{23}\cos\phi+\chi_{13}\sin\phi)\nonumber\\
&+\frac{1}{2}\cos\theta(2\chi_{12}\cos2\phi+(-\chi_{11}+\chi_{22})\sin2\phi),\nonumber\\
\chi_{\hat{\phi}\hat{\phi}}=&\chi_{22}\cos^2\phi-2\chi_{12}\cos\phi\sin\phi+\chi_{11}\sin^2\phi.\nonumber\\
\label{chiall}
\end{align}

It is well known that the gravitational waves have only two physical
freedom in Minkowski background. In 1999, Vega has shown that the
gravitational waves in de Siiter background also share this
property \cite{Vega1999,flanagan2005basics}. When we consider a distant observer in the
wave zone measures gravitational waves from a radiating source at
origin, the metric form will be simple in transverse traceless (TT)
gauge. For this purpose, we need to project out  the non-TT part of
the perturbation, and then the metric can be written
as
\begin{align}
ds^2=&-dt^2+e^{2Ht}\bigg{[}d{R}^2+{R}^2(1+\epsilon
h_+)d\theta^2\nonumber\\
&+{R}^2\sin^2{\theta}(1-\epsilon h_+)d\phi^2+2\epsilon
{R}^2\sin{\theta} h_{\times}d\theta d\phi\bigg{]},\label{desitter3}
\end{align}
with
\begin{align}
&h_+=\frac{1}{2}(\chi_{\hat{\theta}\hat{\theta}}-\chi_{\hat{\phi}\hat{\phi}}),\\
&h_{\times}=\chi_{\hat{\theta}\hat{\phi}}.
\end{align}
To see this, we can use the projection operator
\begin{eqnarray}
 P_i{}^j&=\delta_i{}^j-n_in^j,
 \end{eqnarray}
with $n^i=(\frac{\partial}{\partial{R}})^i$ which is
 normal respect to the Euclidian metric $\delta_{ij}$ and
 raising and lowering index with metric $\delta_{ij}$.
Then the transverse-traceless part of the perturbation can be
calculated as \cite{Shapiro2010}
\begin{eqnarray}
h_{ij}^{TT}&=&(P_i^kP_j^l-\frac{1}{2}P_{ij}P^{kl})h_{kl}\nonumber\\
&=&(\delta_i^k\delta_j^l-\delta_i^kn_jn^l-n_in^k\delta_j^l\nonumber\\
&&+\frac{1}{2}n_in_jn^kn^l+\frac{1}{2}\delta_{ij}n^kn^l)h_{kl}
\end{eqnarray}
Straight forward calculation gives
\begin{eqnarray}
&\chi_{\hat{R}\hat{R}}^{TT}=0, \chi_{\hat{R}\hat{\theta}}^{TT}=0, \chi_{\hat{R}\hat{\phi}}^{TT}=0,\nonumber\\
&h_+=\chi_{\hat{\theta}\hat{\theta}}^{TT}=-\chi_{\hat{\phi}\hat{\phi}}^{TT}=\frac{1}{2}(\chi_{\hat{\theta}\hat{\theta}}-\chi_{\hat{\phi}\hat{\phi}}),\nonumber\\
&h_{\times}=\chi_{\hat{\theta}\hat{\phi}}^{TT}=\chi_{\hat{\theta}\hat{\phi}}.\nonumber
\end{eqnarray}

\section{reexpress solution to linearized Einstein equation with a cosmological constant in Bondi-Sachs framework}

We look for the corresponding Bondi-Sachs coordinate
$(u,r,{\theta},{\phi})$ for the metric (\ref{desitter3}).
 In order to find the coordinate $u$ we have to solve the Eikonal equation
\begin{eqnarray}
g^{ab}\nabla_au\nabla_bu=0.\label{eikonal}
\end{eqnarray}
We solve this equation approximately with the form
\begin{eqnarray}
u=-\frac{1}{H}\ln(H{R}+e^{-tH})+\epsilon
f_1(t,{R},{\theta},{\phi})+o(\epsilon).
\end{eqnarray}
Here the leading order term corresponds to the $u$
coordinate for de Sitter space. Plug the
 metric (\ref{desitter3}) into the equation (\ref{eikonal}), we get
\begin{eqnarray}
e^{-tH}\frac{\partial f_1}{\partial {R}}+\frac{\partial
f_1}{\partial t}=0.
\end{eqnarray}
Since one special Bondi-Sachs coordinate is enough for our usage,
we need only find one special solution to the above equation. For simplicity, we chose the special solution $f_1=0$.

Now we can assume the coordinate transformation from
$(t,{R},{\theta},{\phi})$ to $(u,r,{\theta},{\phi})$ reads as \cite{Bishop16}
\begin{align}
u=&-\frac{1}{H}\ln(H{R}+e^{-tH})+o(\epsilon),\label{eq27}\\
r=&e^{tH}{R}+\epsilon f_2(t,{R},{\theta},{\phi})+o(\epsilon).\label{eq28}
\end{align}
Once again, the leading order term in the above $r$ equation
corresponds to the $r$ coordinate for de Sitter space. Because the
$r$ should be a luminosity distance parameter, within
$(u,r,{\theta},{\phi})$ coordinate we have
\begin{eqnarray}
g_{22}g_{33}-g_{23}^2=r^4\sin^2{\theta},
\end{eqnarray}
where $g_{22}$, $g_{23}$ and $g_{33}$ are the metric coefficients
within $(u,r,{\theta},{\phi})$ coordinate. Direct calculation shows
\begin{eqnarray}
f_2(t,{R},{\theta},{\phi})=0.
\end{eqnarray}

So based on the above coordinate transformation, we can calculate
the metric coefficients within Bondi-Sachs coordinate
$(u,r,{\theta},{\phi})$ as
\begin{align}
g^{{\theta}{\theta}}=&\frac{1}{r^2}-\frac{\epsilon h_{+}}{r^2},\\
g^{{\theta}{\phi}}=&-\frac{\epsilon h_{\times}}{r^2\sin{\theta}},\\
g^{{\phi}{\phi}}=&\frac{1}{r^2\sin^2{\theta}}+\frac{\epsilon
h_{+}}{r^2\sin^2{\theta}}.
\end{align}

Recalling the result (\ref{desitter2}), and using the coordinate
relation between $(t,{R},{\theta},{\phi})$ and
$(u,r,{\theta},{\phi})$ we have
\begin{eqnarray}
h_+=H^2 A+\frac{\dot{A}}{r},\label{anaA}
\end{eqnarray}
with
\begin{align}\label{A}
A=&\bigg{[}(\dot Q_{11}+H(\bar{Q}_{11}-2Q_{11}))(\cos^2\phi\cos^2\theta-\sin^2\phi)\nonumber\\
&+(\dot Q_{22}+H(\bar{Q}_{22}-2Q_{22}))(\sin^2\phi\cos^2\theta-\cos^2\phi)\nonumber\\
&+(\dot Q_{33}+H(\bar{Q}_{33}-2Q_{33}))\sin^2\theta \nonumber\\
&+(\dot Q_{12}+H(\bar{Q}_{12}-2Q_{12}))(1+\cos^2\theta)\sin2\phi \nonumber\\
&-(\dot Q_{13}+H(\bar{Q}_{13}-2Q_{13}))\sin2\theta\cos\phi\nonumber\\
&-(\dot Q_{23}+H(\bar{Q}_{23}-2Q_{23}))\sin2\theta\sin\phi \bigg{]},
\end{align}
and
\begin{eqnarray}
h_{\times}=H^2B+\frac{\dot{B}}{r},\label{anaB}
\end{eqnarray}
with
\begin{align}\label{B}
B=&\bigg{[}2\sin\theta\sin\phi(\dot Q_{13}+H(\bar{Q}_{13}-2Q_{13}))\nonumber\\
&-2\sin\theta\cos\phi (\dot Q_{23}+H(\bar{Q}_{23}-2Q_{23}))\nonumber\\
&+2\cos\theta\cos2\phi(\dot Q_{12}+H(\bar{Q}_{12}-2Q_{12}))\nonumber\\
&-\cos\theta\sin2\phi((\dot Q_{11}-\dot Q_{22})\nonumber\\
&+H(\bar{Q}_{11}-\bar{Q}_{22}-2Q_{11}+2Q_{22}))\bigg{]}.
\end{align}
When $\Lambda=0$ our results for $h_+$ and $h_\times$ recover the results got in the literature (say the Eqs.~(9.24) and (9.25) of Ref.~\cite{Shapiro2010} and the Eq.~(3.72) of Ref.~\cite{Shapiro2010})\footnote{Note that the Eqs.~(9.24) and (9.25) of Ref.~\cite{Shapiro2010} has some typos about terms relating to $Q_{12}$ and $Q_{13}$.}.
So more explicitly we have
\begin{align}
g^{{\theta}{\theta}}=&\frac{1}{r^2}-\frac{\epsilon H^2
A}{r^2}-\frac{\epsilon
\dot{A}}{r^3},\label{New1}\\
g^{{\theta}{\phi}}=&-\frac{\epsilon H^2
B}{r^2\sin{\theta}}-\frac{\epsilon
\dot{B}}{r^3\sin{\theta}},\label{New1b}\\
g^{{\phi}{\phi}}=&\frac{1}{r^2\sin^2{\theta}}+\frac{\epsilon H^2
A}{r^2\sin^2{\theta}}+\frac{\epsilon \dot{A}}{r^3\sin^2{\theta}}.
\end{align}

If the gravitational wave source is axisymmetric, the corresponding spacetime is also axisymmetric. For a axisymmetric source,
the quadruple components satisfy the following relation
\begin{align}
Q_{22}=Q_{11},&Q_{12}=Q_{13}=Q_{23}=0,\\
\bar{Q}_{22}=\bar{Q}_{11},&\bar{Q}_{12}=\bar{Q}_{13}=\bar{Q}_{23}=0.
\end{align}
These relations result in
\begin{align}
A=&\sin^2\theta[\dot{Q}_{33}-\dot{Q}_{11}+H(\bar{Q}_{33}-\bar{Q}_{11}-2(Q_{33}-Q_{11}))],\label{Aaxis}\\
B=&0.
\end{align}

\section{related result based on new Bondi-type boundary condition}

\subsection{Axisymmetric case}
In Refs.~\cite{HeCao15,HeCaoJing16} we have proposed a new Bondi-type
 out going boundary condition for Einstein equation with
 a cosmological constant. For simplicity, we considered only
 axisymmetric spacetime there. Based on axisymmetric assumption,
 the metric can be written as following within
Bondi-Sachs coordinate
\begin{align}
ds^2=&-(Vr^{-1}e^{2\beta}-U^2r^2e^{2\gamma})du^2-2e^{2\beta}dudr\nonumber\\
&-2Ur^2e^{2\gamma}du
d{\theta}+r^2(e^{2\gamma}d{\theta}^2+e^{-2\gamma}\sin^2{\theta}
d{\phi}^2).\label{AaxisBS}
\end{align}
and we have shown that under the outgoing boundary
condition
\begin{eqnarray}
\gamma=\Lambda f(u,{\theta})+\frac{c(u,{\theta})}{r}+\cdots
\end{eqnarray}
the Einstein equation together with a cosmological constant results in
\begin{align}
ds^2=&\bigg{[}\frac{\Lambda}{3}r^2-\frac{c^2\Lambda}{6}-e^{-2\Lambda
f}(1+\frac{1}{2}e^{2\Lambda f}\Lambda c^2+3\Lambda\frac{\partial
f}{\partial{\theta}}\nonumber\\
&+\Lambda\frac{\partial^2
f}{\partial{\theta}^2}-2\Lambda^2(\frac{\partial
f}{\partial{\theta}})^2)+\frac{2M}{r}\bigg{]}du^2\nonumber\\
&-2(1-\frac{c^2}{2r^2})dudr+2\bigg{[}2c\cot{\theta}+\frac{\partial
c}{\partial{\theta}}-2\Lambda
c\frac{\partial f}{\partial{\theta}}\nonumber\\
&+\frac{4c^2\cot{\theta}-U_3e^{2\Lambda f}+2c\frac{\partial
c}{\partial{\theta}}-
4\Lambda c^2\frac{\partial f}{\partial{\theta}}}{r}\bigg{]}dud{\theta}\nonumber\\
&+e^{2\Lambda f}\bigg{[}r^2+2cr+2c^2+(c^3+2C)\frac{1}{r}\bigg{]}d{\theta}^2\nonumber\\
&+e^{-2\Lambda
f}\bigg{[}r^2-2cr+2c^2-(c^3+2C)\frac{1}{r}\bigg{]}\sin^2{\theta}
d{\phi}^2,
\end{align}
up to the order $o(r^{-2})$. Regarding to the meaning of the notations such as $M$ and $C$, we refer readers to our previous papers \cite{HeCao15,HeCaoJing16}. Based on the coordinate transformation Eqs.~(\ref{eq27}) and (\ref{eq28}), the metric (\ref{desitter3}) seems to result in vanishing metric coefficient for $dud\theta$. This is because the  (\ref{desitter3}) solves the Einstein equation up to leading post-Newtonian order. So the back reaction such as the $dud\theta$ term does not show up.

Equivalently, the inverse metric can be calculated as
\begin{eqnarray}
g^{{\theta}{\theta}}=e^{-2\Lambda f}(\frac{1}{r^2}-\frac{2c}{r^3})
=\frac{1}{r^2}-\frac{6H^2f}{r^2}-\frac{2c}{r^3},\label{New11}
\end{eqnarray}
where the relation $\Lambda=3 H^2$ has been used and the higher order terms like $f^2,  cf$  are dropped because we consider weak gravitational radiation which will be used to compare with the solution to the linearized Einstein equation got in above sections.

Comparing Eqs.~$(\ref{New1})$ and $(\ref{New11})$, we can obtain
\begin{eqnarray}
f=\frac{\epsilon A}{6},\ \ \ c=\frac{\epsilon \dot{A}}{2}.
\end{eqnarray}
This solution results in
\begin{eqnarray}
c=3\dot{f},
\end{eqnarray}
which is consistent to the Eq.~(19) of Ref.~\cite{HeCao15} we got previously.

When $\Lambda=0$, Eqs.~$(\ref{New1})$ and $(\ref{New11})$ becomes
\begin{eqnarray}
g^{{\theta}{\theta}}=\frac{1}{r^2}-\frac{\epsilon \dot{A}}{r^3}
\end{eqnarray}
and
\begin{eqnarray}
g^{{\theta}{\theta}}=\frac{1}{r^2}-\frac{2c}{r^3}
\end{eqnarray}
respectively. Combing with Eq.~$(\ref{Aaxis})$ , we can get
\begin{eqnarray}
c=\frac{1}{2}\sin^2\theta[\ddot{Q}_{33}-\ddot{Q}_{11}],
\end{eqnarray}
which is exactly coincide with the result in Minkowski
perturbation \cite{Shapiro2010}.

\subsection{General spacetime case}

In this subsection we extend our previous results to general spacetime situation.
For general spacetime,  the metric within Bondi-Sachs coordinate
can be written as \cite{BonVanMet62,Sac62}
\begin{eqnarray}
ds^2=-(Vr^{-1}e^{2\beta}-r^2h_{AB}U^AU^B)du^2-2e^{2\beta}dudr\nonumber\\
-2r^2h_{AB}U^Adu dx^B+r^2h_{AB}dx^Adx^B,
\end{eqnarray}
where the index $A$ and $B$ run through 2 and 3 which correspond to
${\theta}$ and ${\phi}$ respectively. Using the notations $U^2=U,\ U^3=W\csc{\theta}$ and
\begin{eqnarray}
h_{AB}=\left(
         \begin{array}{cc}
           e^{2\gamma}\cosh 2\delta & \sinh 2\delta\sin{\theta} \\
           \sinh 2\delta\sin{\theta} & \ \ \ e^{-2\gamma}\cosh2\delta\sin^2{\theta} \\
         \end{array}
       \right),
\end{eqnarray}
where $\beta$, $\gamma$, $\delta$, $V$, $U$, and $W$ are six
functions of $u$, $r$ and points on the 2-sphere which are
parameterized by $\theta$ and $\phi$. It can be easily
checked that
\begin{eqnarray}
\det(h_{AB})=\sin^2{\theta},
\end{eqnarray}
which guarantees that $r$ is a luminosity distance parameter. When $\delta=W=0$, the line element reduces to
the axisymmetric case (\ref{AaxisBS}).

Regarding to the boundary condition, we need to specify both $\gamma$ and $\delta$ for general spacetime. Corresponding to the new Bondi-type out going boundary condition, the specific form reads as
\begin{align}
\gamma=&\Lambda
f(u,\theta,\phi)+\frac{c(u,{\theta},{\phi})}{r}+\cdot\cdot\\
\delta=&\Lambda\tilde{f}(u,{\theta},{\phi})+\frac{\tilde{c}(u,{\theta},{\phi})}{r}+\cdots
\end{align}
Based on these boundary conditions, we can solve Einstein equation together with a cosmological constant to get
\begin{align}
g^{{\theta}{\theta}}=&\frac{e^{-2\Lambda f}\cosh 2\Lambda
\tilde{f}}{r^2}\nonumber\\
&+\frac{e^{-2\Lambda f}(-2c\cosh
2\Lambda\tilde{f}+2\tilde{c}\sinh2\Lambda\tilde{f})}{r^3}\nonumber\\
\approx&\frac{1}{r^2}-\frac{6H^2f}{r^2}-\frac{2c}{r^3},\\
g^{{\theta}{\phi}}=&-\frac{\sinh
2\Lambda\tilde{f}}{r^2\sin{\theta}}-\frac{2\tilde{c}\cosh2\Lambda\tilde{f}}{r^3\sin{\theta}}\nonumber\\
\approx&-\frac{6H^2
\tilde{f}}{r^2\sin{\theta}}-\frac{2\tilde{c}}{r^3\sin{\theta}}.
\end{align}
Here ``$\approx$" means we have already dropped the nonlinear terms
respect to $f$, $\tilde{f}$, $c$ and $\tilde{c}$ based on the
assumption that the gravitational radiation is weak. The Einstein
equation together with a cosmological constant will give us a
relation
\begin{align}
c=3\dot{f},\ \ \ \tilde{c}=3\dot{\tilde{f}}.\label{relation3}
\end{align}
Compared to (\ref{New1}) and (\ref{New1b}) we have
\begin{align}
f=\frac{\epsilon A}{6},&\ \ \ c=\frac{\epsilon \dot{A}}{2},\\
\tilde{f}=\frac{\epsilon B}{6},&\ \ \ \tilde{c}=\frac{\epsilon
\dot{B}}{2}.
\end{align}
Explicitly, this result is consistent to the above Eqs.(\ref{relation3}).

\section{Summary and discussion}

After the direct detection of gravitational wave including events GW150914, GW151226 and GW170104, the exciting gravitational wave astronomy era is coming.  We expect to get some hints about  the mysterious problems such as dark energy through gravitational wave detection. Considering the fact that our universe is expanding, we have to understand theoretically well how the cosmological constant affect the behavior of the gravitational wave.

Although the cosmological constant of our Universe is tiny,  the
authors in Ref.~\cite{Ashtekar16} theoretically found that it may change
the behavior of the gravitational wave strongly. Noting that
Bondi-Sachs framework is  a powerful tool to treat gravitational wave problem \cite{BonVanMet62,Caohe13}, we have used it to investigate the asymptotical behavior of spacetime \cite{HeCao15,HeCaoJing16}. It was found that the original out going boundary condition proposed by Bondi and his coworkers seems break down if we consider the cosmological constant effect. Then we investigate gravitational radiation in the presence of a cosmological constant by imposed a new Bondi-type out going boundary condition which implies that the tiny cosmological constant does affect the gravitational wave perturbably.
On the other hand, in Refs.~\cite{Ashtekar2015II,Ashtekar2015III,Date2015}, the authors
solved the linearized Einstein equation together with a cosmological
constant. The solutions describe the behavior of gravitational wave produced by weak sources and the quadrupole formula was given.

An interesting question is how to interpret  the quantities in asymptotic null structure in BS framework in terms of the information of the source in the perturbation approach. For this purpose, it is needed to find the
relation between BS coordinate used in BS framework and generalized harmonic
coordinate used in linearized perturbation method. Coleman
investigated this problem by solving the harmonic gauge condition in the case without a cosmological constant,
but it is too complicated to solve the generalized harmonic
condition in the case with a positive cosmological constant.
In this paper, we transform the linearized solution to the transverse-traceless gauge firstly and  then to Bondi-Sachs gauge by
solving the eikonal equation and luminosity distance condition. Finally we found the explicit expression of BS quantities in terms of the quadruple
(including mass quadrupole and pressure quadrupole) of the source.

By the way, it should be noted that  the perturbation begins with the term of $\sqrt{\Lambda}$ quantitatively
by the linearized way. While the results resulted from our new boundary condition begin
with the term of $\Lambda$. Our results in this paper implies that the term of $\sqrt{\Lambda}$ is due to the gauge effect. And more, the behavior predicted by the
linearized solution is exactly consistent to the results resulted from our new boundary condition.  If we take the cosmological constant $\Lambda \rightarrow 0$ , our result is reduced to the Coleman's one.

Now based on Eqs.~(\ref{anaA}) and (\ref{anaB}) we can estimate how far away
when a gravitational wave source is, the cosmological constant may affect the detection of the gravitational wave. If only \begin{align}
r\gtrsim\frac{\omega}{\Lambda}
\end{align}
where $\omega$ is the frequency of the gravitational wave, the cosmological constant may affect the detection of the gravitational wave.
The observations \cite{Pla13} show that the amplitude
of the cosmological constant $\Lambda$ is about $10^{-52}$m${}^{-2}$
in geometric units ($c=G=1$). If we consider LIGO type detectors, the frequency of the detected
gravitational wave is about 100Hz which corresponds to $10^6$m. So only when the source
is farther than $10^{58}$m ($\approx10^{42}$ light years), the cosmological constant may affect
the detection of LIGO type detectors. So we do not expect the cosmological constant may affect the
detection of LIGO type detectors. Similarly, even for pulsar timing method, we do not expect the
cosmological constant may affect the detection of gravitational wave. For the gravitational wave
 with frequency $10^{-16}$Hz which corresponds to the age of our Universe, if the sources
 are $10^{40}$m ($\approx10^{24}$ light years) away, the cosmological constant may affect
 the gravitational wave detection. But note that the time of $10^{24}$ years is much larger
 than the age of our Universe, we do not expect the cosmological constant may affect the gravitational
 wave detection done by our humanbeing. Of course, the above estimate is based on the current perturbation result.
 Regarding to strong gravitational wave sources, more analysis is needed.

\section*{Acknowledgments}
We thank XiongJun Fang, Xiaoning Wu and Songbai Chen for many useful
discussions and comments on the manuscript. Z. Cao was supported
by the NSFC (No.~11375260, No.~11622546, and No.~11690023). X. He was supported
by the NSFC (No.~11401199) and the Research Foundation of Education Bureau of Hunan Province, China
(No.~14C0254). J. Jing was was supported
by the NSFC (No.~11475061).

\bibliography{refs}

\end{document}